\documentstyle[11pt,newpasp,twoside,epsf]{article}

\def\edcomment#1{\iffalse\marginpar{\raggedright\sl#1\/}\else\relax\fi}
\reversemarginpar

\begin{document}

\title{Chandra Analysis of Abell 496}
\author{Renato Dupke}
\affil{University of Michigan, Ann Arbor}

\author{Raymond E.~White III}
\affil{University of Alabama, Tuscaloosa}

\begin{abstract}
We present preliminary results of imaging and spectral analysis of the 
cluster A496 with the S3 chip
on-board the Chandra satellite. We confirm the presence of a cooling flow 
and radial gradients in elemental
abundances. The cluster's core has clear signs of substructures 
(sharp edges, cavities), similar to those
recently found in other cooling flow clusters. The O/Fe ratio distribution
indicates that the gas in the core of the cluster is significantly 
enriched by SN Ia ejecta.
\end{abstract}

\section{Introduction}

The detection of emission lines in the X-ray spectra of clusters shows 
the presence of heavy elements in the ICM. This fact alone shows that the 
ICM cannot be completely primordial. A significant part of the gas has 
been contaminated by supernovae enriched material, most likely ejected from cluster galaxies.
Competing metal enrichment mechanisms, such as protogalactic winds, ram 
pressure stripping, etc., are powered by different SN Types. Therefore, their 
relative importance can be assessed if the spatial distribution of SN Type 
contamination is known.

The latter can be obtained through spatially resolved X-ray spectroscopy 
(e.g. Mushotzky et al. 1996). Since SN Ia and II explosions produce different 
amounts of different elements, the SN Ia/II contamination fraction in the ICM 
can be determined through the X-ray measurements of individual elemental 
abundances (e.g. O, Si, S, Fe, Ni) and their ratios. For example, SNe Ia 
produce roughly five times less Si than Fe (by mass) and SNe II produce in 
average $\sim$ 30\% more Si than Fe. X-ray spectroscopic determinations of Si and Fe 
abundances can then be used to determine the fraction of the elements in the 
ICM that came from SN Ia and II, and also the spatial variation of this fraction
within the cluster.

Using arguments based on the energetics of the intracluster gas, White (1991) 
has shown that protogalactic winds significantly pollute the ICM with heavy 
elements. Early Einstein FPCS spectroscopy (Canizares et al. 1982) and more recent
ASCA spectroscopy (Mushotzky \& Loewenstein 1997; Mushotzky et al. 1996) showed 
that global intracluster metal abundance ratios are consistent with ejecta from 
Type II supernovae, which supports the protogalactic wind model for ICM metal 
enrichment. 

However, more recently, ASCA spatially resolved spectra of some nearby cool (T $<$ 5keV) 
clusters and groups indicates a significant SN Ia contamination within the 
cluster's central regions (e.g. Dupke \& Arnaud 2001; Dupke \& White 2000; 
Finoguenov et al. 2000; Allen et al. 2000), which suggests that a 
combination of ram-pressure stripping \& SN Ia secondary winds may be also 
effective in injecting metals into the ICM.

\begin{figure}
\caption{X-ray spectrum of the O~K$_{\alpha}$(top) ($>$2$^{\prime}$ from the center) and FeL line complex (bottom) in the central 30$^{\prime\prime}$  (Fig 4 right)}
\vspace{2.5truein}
\plotfiddle{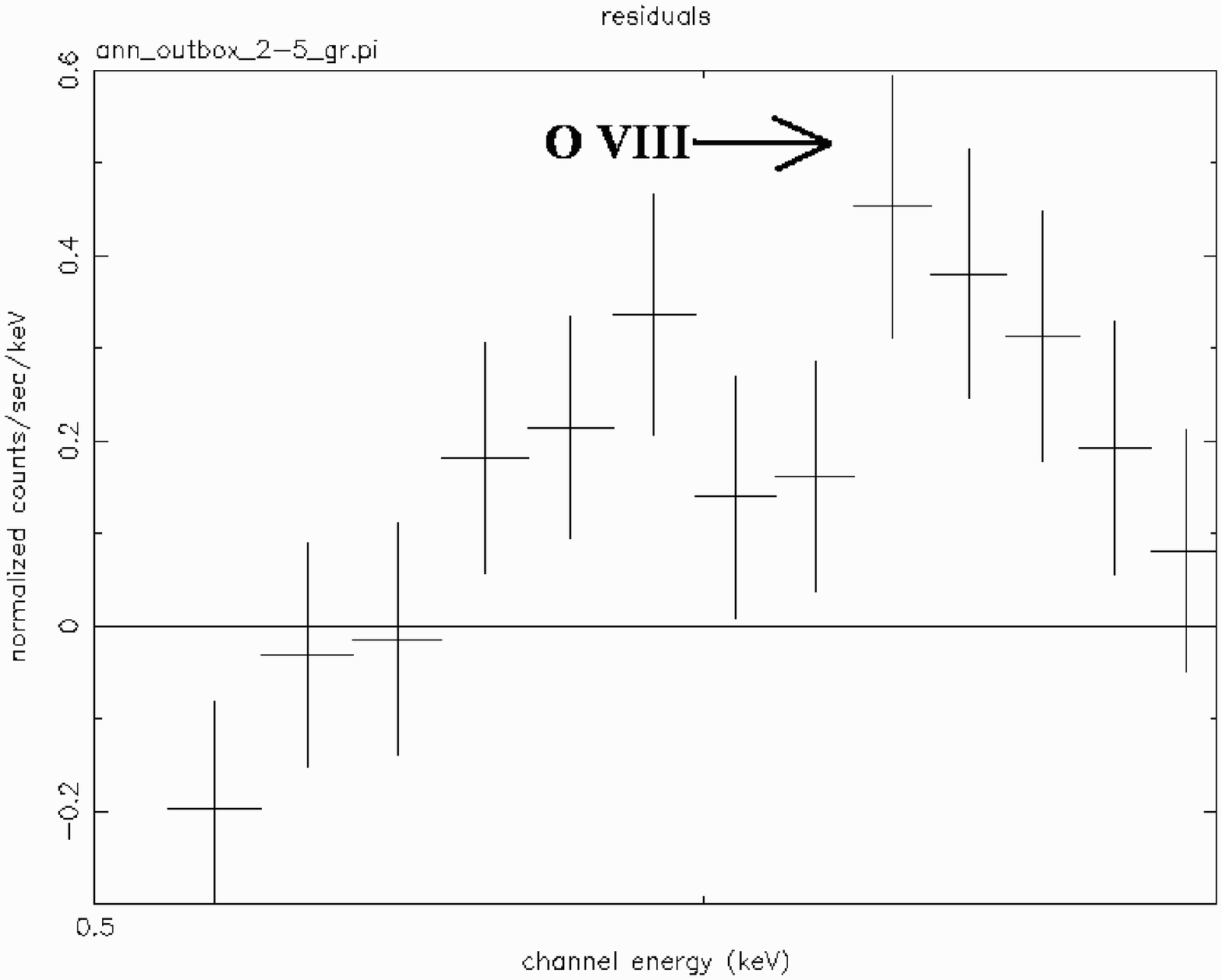}{0.1in}{0.0}{40}{20}{-130.0}{47.0}
\plotfiddle{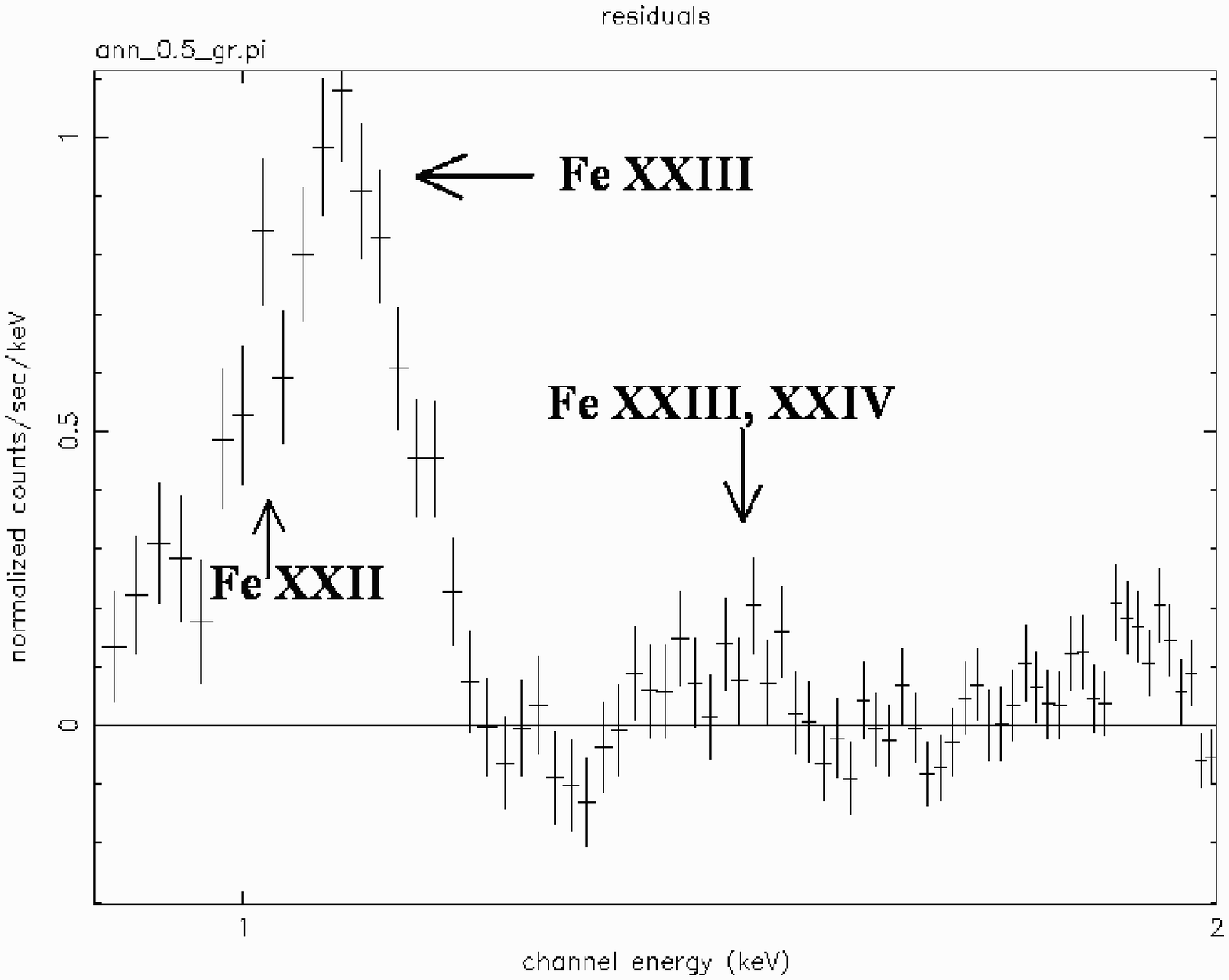}{0.1in}{0.0}{40}{20}{-130.0}{-50.0}
\end{figure}

Abell 496 is a typical, bright, nearby (z$\sim$0.03), well-behaved cooling 
flow cluster. In a Ginga + Einstein analysis, White et al. (1994) detected 
a central abundance enhancement in this cluster. This central abundance 
enhancement was corroborated by ASCA (e.g. Dupke \& White 2000), SAX 
(Irwin \& Bregman 2001) and XMM (Tamura et al 2001). Furthermore, 
Dupke \& White (2000) also detected radial gradients of 
abundance ratios, which showed that the gas in the central regions had been 
significantly contaminated by SN Ia ejecta. Our Chandra observations of the 
central regions of A496 allow us to expand the
measurements of abundance and abundance ratios to smaller spatial scales 
(down to $<$1$^{\prime}$) and also to obtain well-constrained abundances of elements that have high 
SN-Type discriminatory power such as Oxygen.

\section{Data Reduction}

We used Ciao 2.1 with caldb 2.3. The data was screened as prescribed in the
Threads page\footnote{asc.harvard.edu/ciao2.1/documents\_threads.html}, 
complemented by the extended source reduction procedures by J. 
McDowell\footnote{asc.harvard.edu/ciao/wrkshp\_ps.html} with the response 
tools developed 
by A. Vikhlinin\footnote{http://hea-www.harvard.edu/\~jcm/asc/dist/av/av103.tar}. 
Approximately 70\% of the data was seriously affected by background 
flares and was not used in this analysis. Both 
blank-sky\footnote{hea-www.harvard.edu/\~maxim/axaf/acisbg/} and local 
(from the S1 chip) 
backgrounds were tried. The latter  was found to minimize
flare contamination effects better. Thermal emission models mekal \& Vmekal (individual 
abundances are free to vary) were used within XSPEC 11 to fit X-ray spectra. 
An isobaric cooling flow model Cflow, was used for the spectral fittings of 
the inner regions where the minimum temperature was either frozen at 0.1 keV 
or let free to vary. Some spurious spectral features were seen around 2 keV 
and are probably related to background correction uncertainties. Here we only show here 
the results of the well-constrained abundance 
distributions of O and Fe (Fig 2).

\begin{figure}
\caption{Radial distribution of temperature and abundances. All errors 
are 90\% confidence. Dashed lines indicate the 90\% confidence limits
from ASCA (Dupke \& White 2000). Solid black lines show fittings results for a Vmekal + Cflow model.
 Brighter lines show the fitting results for a simple Mekal model.}
\vspace{3.0truein}
\plotfiddle{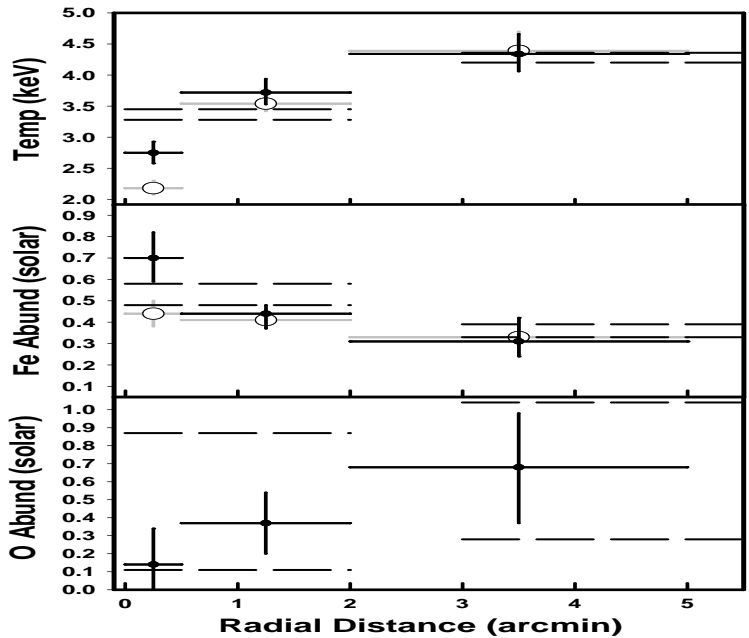}{0.1in}{0.0}{65}{35}{-180.0}{-35.0}
\end{figure}

\section{Results and Discussion}
Fig 2 shows the radial distribution of temperature and abundances in the 
central 5$^{\prime}$ of the cluster. A negative temperature gradient is clearly seen, consistent with the presence 
of a cooling flow. There are indications that the minimum gas temperature of 
the cool gas component is $\sim$1 keV. The spectral 
fits when the elemental abundances were let free to vary (Vmekal) were 
significantly better than those where the abundances are tied to the solar 
ratios (large circles with brighter errorbars). This is most likely due to abundance ratios being 
strongly non-solar within the central regions. 
The best-fit values of this model (added to a cooling flow component in the 
central regions) are indicated by the solid plots.
\begin{figure}
\caption{O/Fe ratio and SN Ia Fe Mass Fraction. The errors 
are 1-$\sigma$. Dashed lines indicate the 90\% confidence limits
from ASCA (Dupke \& White 2000) for comparison.}
\vspace{3.5truein}
\plotfiddle{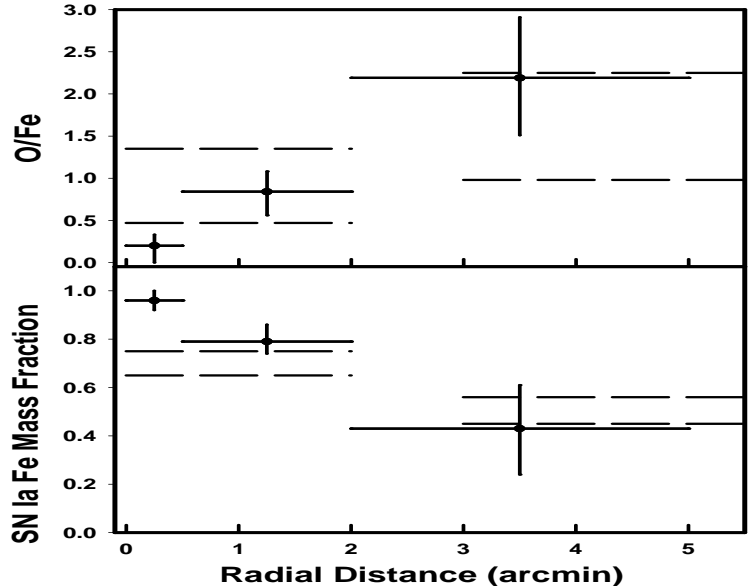}{0.1in}{0.0}{65}{40}{-180.0}{-35.0}
\end{figure}

Previous results obtained with ASCA by the authors are displayed by the dashed lines 
(90\% confidence limits) for the inner (0$^{\prime}$-2$^{\prime}$) 
and outer (3$^{\prime}$-12$^{\prime}$) regions 
obtained with the same spectral models as those used here. There was a 
significant improvement ($>$99.9\% confidence in the F-test) when the minimum 
temperature (T$_{min}$) in the cooling flow model was let free to vary. 
The best-fit T$_{min}$ was found to be 1$\pm$0.2 keV. The Fe abundance gradient is confirmed. 
The gradient is apparently present 
even at very small scales (within 2$^{\prime}$), where the Fe abundance in the very center is found to be 
as high as $\approx$0.7 solar. The Oxygen abundance shows a significant radial enhancement,
rising from $\sim$0.15 solar in the central 30$^{\prime\prime}$ to $\sim$0.7 solar at $>$2$^{\prime}$,
which is consistent with previous ASCA results, but marginally consistent with recent XMM analysis of 
this cluster (Tamura et a. 2001). 

Fig. 3 shows the radial distribution of the O/Fe ratio and the correspondent 
SN Ia contamination (using SN yields of Nomoto et al. 1997a, b). The errors 
are 1-$\sigma$, and the confidence limits showed (dashed) for the bottom plot 
correspond to the SN Ia Fe mass fraction determined with ASCA from an {\it ensemble}
of abundance ratios including O, Si, Ne, Fe and Ni). 
\begin{figure}
\caption{Left - SN Ia contamination in A496 -- Taken as FeL/O 
image with continuum subtracted. 
Right - S3 smoothed image of the central arcmin of A496}
\vspace{3.0truein}
\plotfiddle{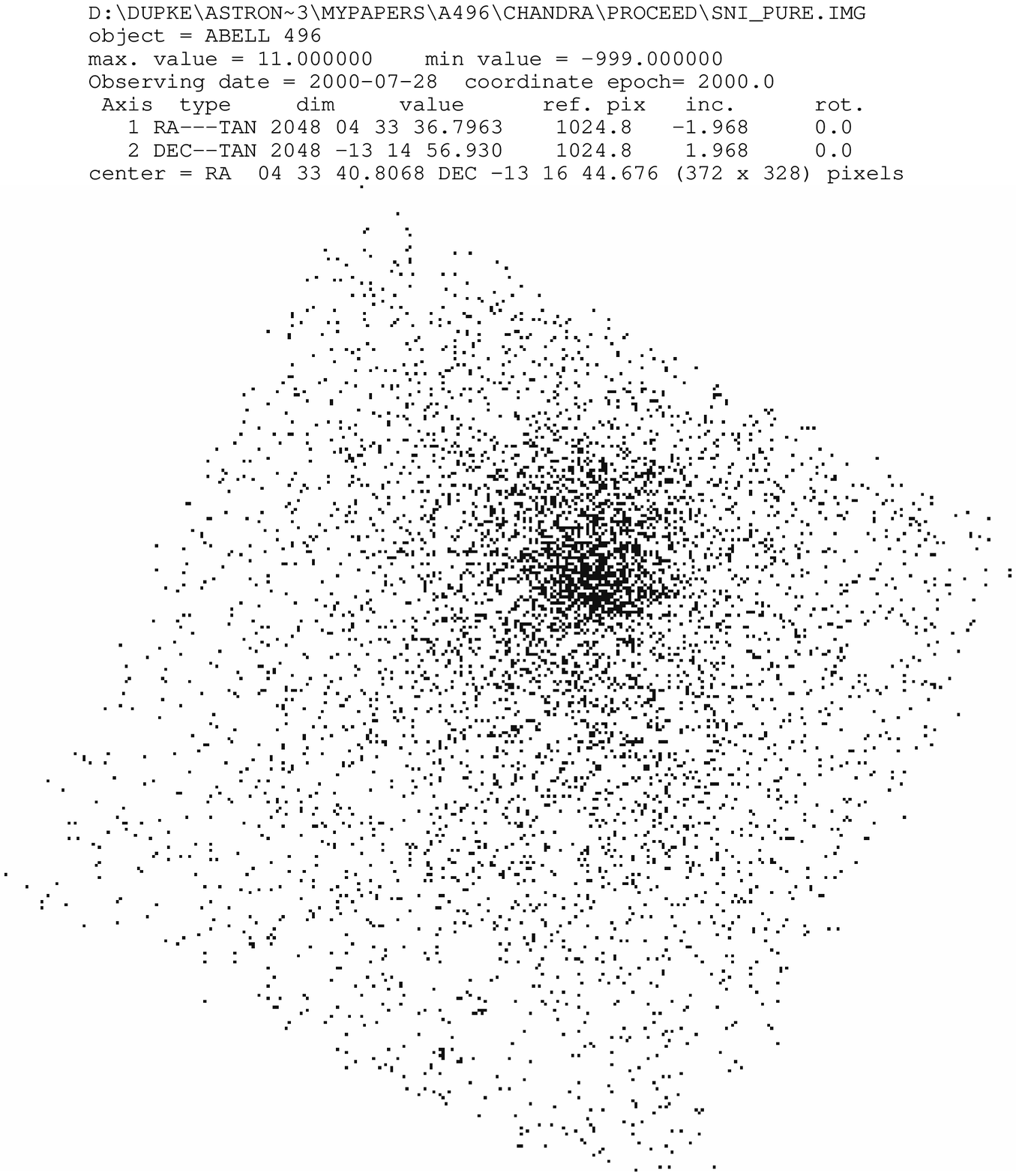}{0.1in}{0.0}{40}{40}{-200.0}{-80.0}
\plotfiddle{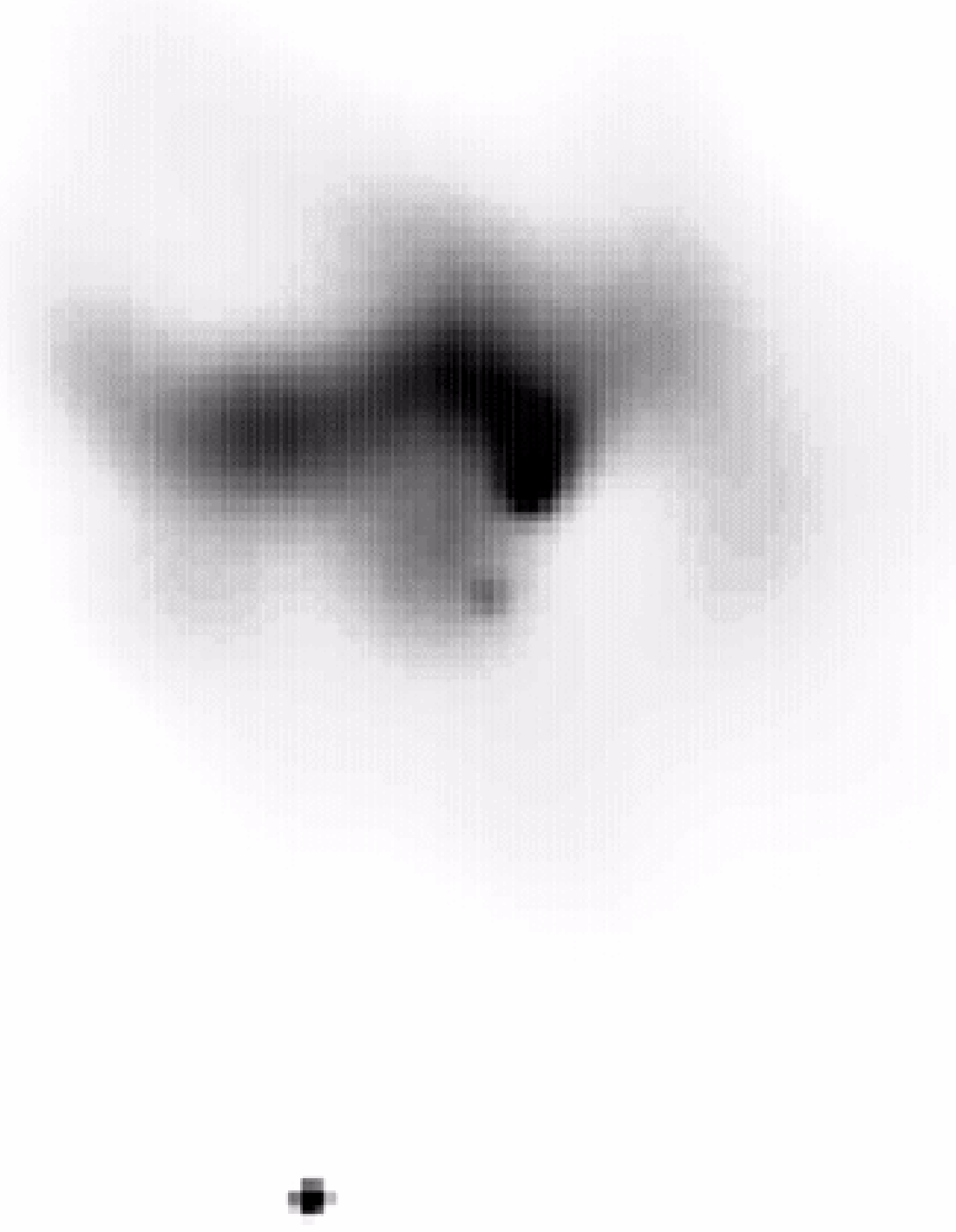}{0.1in}{0.0}{40}{40}{5.0}{-70.0}
\end{figure}

The nature of the gradients in the radial distribution O (positive) and Fe 
(negative) implies on a positive gradient of the O/Fe abundance ratio. This 
gradient can be associated to a gradient of SN Type responsible for the gas 
enrichment. For comparison the O/Fe ratio for "pure" SNe Ia is 0.03 and for
SNe II is 3.8. The O/Fe ratio in the ICM of A496 varies from $\sim$0.2 in the 
very center (implying that the $>$ 90 \% of the metals come from SN Ia) to 
$\sim$2.2 at $>$2$^{\prime}$, where an even mix of SN Ia and II contribution is implied.
Fig.4 (left) shows a ``SN Ia'' map of the cluster. The map is obtained by dividing
images taken within the FeL energy range by the O energy range image. The continuum is
subtracted by selecting similar images within nearby frequencies. It can be seen that
SN Ia distribution is enhanced towards the center.

The results above are consistent with the double wind scenario for ICM metal 
enrichment proposed by Dupke \& White (2000). A vigorous protogalactic winds 
that injects metals produced by SN II uniformly into the ICM followed by a 
secondary, less energetic SN Ia dominated wind, which is partially suppressed 
at the cluster's center due to the high ambient intracluster gas density at 
the bottom of the clusters gravitational potential.

Multiple X-ray filaments can be seen in the central region of A496 (Fig4, right). 
In particular a bright filament is detected extending from the nucleus 
to the East. The X-ray filament extension is $\sim$20$^{\prime \prime}$
($\sim$19 $h_{50}^{-1}$ kpc).
Radio emission (VLA 1.4 GHz) of the region around the cD, MCG 02-12-039 is 
slightly extended towards the direction of the X-ray filament. There are 
optical emission filaments in the central regions of A496 (Fabian et al. 1981). 
However, the optical data is not spatially resolved enough for a fine comparison.
The X-ray central structures are seen mostly at the medium energy range 
(0.8-1.5 keV). A more detailed discussion about the morphological structures 
in A496 will be made in a future work (Dupke \& White 2001).

\end{document}